\documentclass{aa}  

\usepackage{graphicx}
\usepackage{txfonts}
\usepackage{natbib}
\usepackage{longtable}
\usepackage{lscape} 
\usepackage{graphicx}
\usepackage{url}
\usepackage{amssymb}

\begin{document}

   \title{Messier 35 (NGC~2168) DANCe}

   \subtitle{I. Membership, proper motions and multi-wavelength photometry}

   \author{H. Bouy \inst{1}
          \and E. Bertin\inst{2}
          \and D. Barrado\inst{1}          
          \and L.M. Sarro\inst{3}
          \and J. Olivares\inst{3,4,5}
          \and E. Moraux\inst{4,5}
          \and J. Bouvier\inst{4,5}
          \and J.-C. Cuillandre\inst{6}
          \and \'A. Ribas\inst{1,7,8}
          \and Y. Beletsky\inst{9}
}

   \institute{Centro de Astrobiolog\'\i a, depto de Astro\'\i sica, INTA-CSIC, PO BOX 78, E-28691, ESAC Campus, Villanueva de la Ca\~nada, Madrid, Spain\\
         \email{hbouy@cab.inta-csic.es} \\
         \and
         Institut d'Astrophysique de Paris, CNRS UMR 7095 and UPMC, 98bis bd Arago, F-75014 Paris, France  \\
         \and
         Dpt. de Inteligencia Artificial , UNED, Juan del Rosal, 16, 28040 Madrid, Spain \\
         \and
         Univ. Grenoble Alpes, IPAG, F-38000 Grenoble, France 
         \and 
         CNRS, IPAG, F-38000 Grenoble, France 
         \and
         Canada-France-Hawaii Telescope Corporation, 65-1238 Mamalahoa Highway, Kamuela, HI96743, USA \\
         \and
         European Space Astronomy Centre (ESA), P.O. Box, 78, 28691 Villanueva de la Ca\~{n}ada, Madrid, Spain \\
   		\and ISDEFE - ESAC, P.O. Box, 78, 28691 Villanueva de la Ca\~{n}ada, Madrid, Spain \\
   		\and	
         Las Campanas Observatory, Carnegie Institution of Washington, Colina el Pino, 601 Casilla, La Serena, Chile\\
             }

   \date{Received ; accepted 18/01/2015 }

 
  \abstract
   {Messier 35 (NGC~2168) is an important young nearby cluster. Its age, richness and relative proximity make it a privileged target for stellar evolution studies. The Kepler K2 mission recently observed it and provided high accuracy photometric time series of a large number of sources in this area of the sky. Identifying the cluster's members is therefore of high importance to optimize the interpretation and analysis of the Kepler K2 data.}
   {We aim at identifying the cluster's members by deriving membership probabilities for the sources within 1\degr\, of the cluster's center, going further away than equivalent previous studies.}
   {We measure accurate proper motions and multi-wavelength (optical and near-infrared)  photometry using ground based archival images of the cluster. We use these measurements to compute membership probabilities. The list of candidate members from \citet{2001ApJ...546.1006B} is used as training set to identify the cluster's locus in a multi-dimensional space made of proper motions, luminosities and colors.}
   {The final catalog includes 338\,892 sources with multi-wavelength  photometry. Approximately half (194\,452) were detected at more than two epochs and we measured their proper motion and used it to derive membership probability. A total of 4\,349 candidate members with membership probabilities greater than 50\% are found in this sample in the luminosity range between 10 and 22~mag.  The slow proper motion of the cluster and the overlap of its sequence with the field and background sequences in almost all color-magnitude and color-color diagrams complicate the analysis and the contamination level is expected to be significant. Our study nevertheless provides a coherent and quantitative membership analysis of Messier 35 based on a large fraction of the best ground-based data sets obtained over the past 18 years. As such, it represents a valuable input for follow-up studies using in particular the Kepler K2 photometric time series.}
   {}


   \maketitle
%

\section{Introduction}

Different galactic environments, it is assumed,  produce distinct
 populations during the star formation episodes.
From very loose and young  associations  such as Taurus-Auriga, 
doomed to be dispersed  very soon in the galactic gravitational well,
 to the very old halo population  in massive globular clusters, there is a 
large gamut in ages, compactness, metallicity, internal dynamics, 
total mass, spatial 
distribution and distribution of the individual masses,
 perhaps indicating differences in the
 Initial Mass Functions (IMF). It is clear that a significant improvement in the
completeness of the census of different associations provide key information to 
understand star formation and evolution.

In this context, the open cluster Messier 35 (NGC~2168, hereafter M35) is very interesting in several aspects. First because it is a 
 quasi-twin of the Pleiades cluster, with a similar age \citep{Vidal1973.M35.photometry} but a different metallicity. Second, the Kepler K2 mission \citep{2014PASP..126..398H} has just provided exquisite photometric time series, so that a detailed distribution of  the rotational periods for different masses will be available \citep[e.g][]{Meibom2009.M35.Prot}. Moreover, M35  is not so crowded that confusion is a major problem and its size (about  1 sq. deg) is very suitable
 for follow-up multi-object spectroscopy \citep{Barrado2001.M35.lithiumFe}.

Several studies have provided estimates of other key cluster parameters such as its age, its census and its metallicity. Table~\ref{prev_studies} gives an overview of the cluster's properties found in the literature in the past fifteen years.
However, there is no a clear consensus for most of them.
\citet{1999MNRAS.306..361S} derived a distance modulus of 
$(m-M)_0=9.60\pm0.10$ corresponding to 832~pc, a reddening of
$E(B-V)=0.255\pm0.024$~mag and an age of
200$^{+200}_{-100}$~Myr. \citet{Kalirai2003.M35.photometry} pushed the cluster a little bit further, to
$912^{+70}_{-65}$~pc and derived an isochrone fitting age of 180~Myr.
More recently, \citet{McNamara2011.M35.propermotion} obtained a distance of $732\pm145$~pc from the internal proper motions. This dynamical distance and the radial velocity dispersion obtained by \citet{Geller2010.M35.RV} imply an age of 133~Myr, according to the same authors.
An age  determination based on a completely different technique
 --gyrochronology-- was performed by \citet{Meibom2009.M35.Prot},
 yielding a value between  134~Myr and 161~Myr.
\citet{Reimers1988.M35.WD}, by using the white dwarf cooling age, derived an age in the range 70-100~Myr, whereas 
\citet{Barrado2001.M35.WDage.metal.li.IMF}, with the same technique, estimated 180~Myr.
The rotation pattern suggests an age slightly older than the Pleiades (175~Myr compared to 125 Myr). The same conclusion is reached when the lithium abundances are compared, since the lithium dispersion, which is itself related to rotation for a given stellar mass,  is no longer obvious for M35 \citep{Barrado2001.M35.lithiumFe} while it is very significant in the case of the Pleiades \citep[see, for instance][]{1993AJ....106.1059S}.

\begin{table*}
\caption{Overview of previous studies of M35\label{prev_studies}}
\begin{tabular}{lcccc}\hline\hline
Reference   & Distance & Age & $E(B-V)$ & $[Fe/H]$  \\
            &  (pc)    & (Myr) &  (mag) &        \\
\hline
\citet{Reimers1988.M35.WD} & & 70--100 & & \\
\citet{1999MNRAS.306..361S} & 832$\pm$39 & 200$^{+200}_{-100}$ & 0.255$\pm$0.024 &  \\
\citet{Barrado2001.M35.WDage.metal.li.IMF} & & 180 & & \\
\citet{Barrado2001.M35.lithiumFe} & & $>$125 & & $-0.21\pm0.10$ \\
\citet{Kalirai2003.M35.photometry} & 912$^{+70}_{-65}$ & 180 &  &  \\
\citet{Steinhauer2004.M35.lithiumF} & & & & $-0.143\pm0.014$ \\
\citet{Meibom2009.M35.Prot} &  & 134--161 & & \\
\citet{Geller2010.M35.RV} & & 133 & & \\
\citet{McNamara2011.M35.propermotion} & 732$\pm$145 & & & \\
\hline
\end{tabular}
\end{table*}

Regarding chemical composition, \citet{Steinhauer2004.M35.lithiumF} quoted an unpublished work with a metallicity $[Fe/H]=-0.143\pm0.014$, very similar to the value obtained by our group ($[Fe/H]=-0.21\pm0.10$)
using high spectral resolution data and spectral synthesis, 
 \citep{Barrado2001.M35.lithiumFe}.

The cluster is also relatively massive: \citet{Leonard1989.M35.Mass} derived a central total mass in the range 1600-3200 M$_\odot$ using dynamical models. Our own estimate, for the central of 27.5$\times$27.5 square arcmin, is $\sim1600$
  M$_\odot$, which allows a exquisite determination of a complete stellar mass function \citep[][so far the deepest for M35]{Barrado2001.M35.IMF} which is totally compatible with the IMF of a much younger and un-evolved cluster such as Collinder 69 
\citep{Barrado2005.C69.IMFcomplete,Bayo2011.C69.IMF}, and suggests that M35 includes several tens of potential substellar members.

The goal of this paper is to complete the census reaching well inside the
substellar domain and including the cluster outskirts up to a radius of 1\degr, providing multi-epoch, multi-band photometry from public archives 
and from our own recent observations.


\section{Datasets}
We searched the CFHT, NOAO, Subaru, Kiso, SDSS \citep[DR9, ][]{2012ApJS..203...21A} and ING public archives for wide field images of the cluster. Table~\ref{table_obs} gives an overview of the dataset found in these archives and in our own private archives.

\begin{table*}
\caption{Instruments used in this study\label{table_obs}}
\scriptsize
\begin{tabular}{lcccccccc}\hline\hline
Observatory   & Instrument        & Filters              & Platescale     & Chip layout   & Chip size  & Field of view &        Epoch(s)      & Ref. \\
                      &                          &                         & [pixel$^{-1}$] &                      &                  &                     &                           &         \\
\hline
CFHT              & UH~8K             & R, I                    & 0\farcs205 & 4$\times$2         &2k$\times$4k  & 29\arcmin$\times$29\arcmin & 1996                                &  1 \\
CFHT              & CFHT~12K        & B,V,R,I,z            & 0\farcs201 & 6$\times$2         & 2k$\times$4k & 42\arcmin$\times$28\arcmin & 1999, 2000, 2001, 2002  & 2  \\
CFHT              & MegaCam         & g,r                    & 0\farcs187 & 4$\times$9         & 2k$\times$4k & 1\degr$\times$1\degr & 2003, 2005, 2011  &  3 \\
CFHT              & WIRCam           & H2                    & 0\farcs304 & 2$\times$2         &  2k$\times$2k & 20\arcmin$\times$20\arcmin & 2008, 2009, 2010, 2011   &  4 \\
Subaru            & Suprime-Cam  & VR-Broad          & 0\farcs200 & 5$\times$2         & 2k$\times$4k &  34\arcmin$\times$27\arcmin & 2005     &  5 \\
INT                 & WFC                 & g, H$\alpha$, He~I, i, I, r, U, V   & 0\farcs333 & 3$\times$1+1    & 2k$\times$4k &  34\arcmin$\times$34\arcmin & 2000, 2002, 2003, 2005, 2008, 2009, 2010  &  6  \\ 
KPNO (Mayall) & NEWFIRM         & J, H, Ks              & 0\farcs400 & 2$\times$2         & 2k$\times$2k &  28\arcmin$\times$28\arcmin & 2008               & 7 \\
KPNO (Mayall) & MOSAIC1         & VR-broad          & 0\farcs26   & 4$\times$2 & 2k$\times$4k & 36\arcmin$\times$36\arcmin & 2006, 2009, 2010, 2011, 2012 & 8 \\
KISO               & 2KCCD             & B,V,R,I                & 1\farcs50  & 1$\times$1 & 2k$\times$2k & 50\arcmin$\times$50\arcmin & 2002, 2003 & 9 \\
SDSS DR9 & SDSS camera & $u,g,r,i,z$ & 0\farcs396 & 6$\times$5 & 2k$\times$2k & 3\degr & 2006 \\
\hline
\end{tabular}

References: (1) \citet{1995AAS...187.7305M}, (2) \citet{2000SPIE.4008.1010C}, (3) \citet{2003SPIE.4841...72B}, (4) \citet{2004SPIE.5492..978P}, (5) \citet{2002PASJ...54..833M}, (6) \citet{1998IEEES..16...20I}, (7) \citet{2003SPIE.4841..525A}, (8) \citet{2000SPIE.3965...80W}, (9) \citet{2001PNAOJ...6...41I}, (10) \citet{2012ApJS..203...21A} 
\end{table*}

In each case, the raw data and associated calibration frames were downloaded. The individual raw images were processed using an updated version of \emph{Alambic} \citep{Alambic}, a software suite developed and optimized for the processing of large multi-CCD imagers, and adapted for the instruments used here. \emph{Alambic} includes standard processing procedures such as overscan, bias and dark subtraction for each individual readout ports of each CCD, flat-field correction, bad pixel masking, CCD-to-CCD gain harmonization in the case of multi-chip instruments, fringing correction (when needed for the reddest filters), de-stripping, background subtraction and non-linearity correction in the case of near-infrared detectors. 

Aperture, PSF and model photometry were extracted from the individual images using {\sc SExtractor} \citep{1996A&AS..117..393B} and {\sc PSFEx} \citep{PSFEx}. The individual catalogs were then registered and aligned on the same photometric scale using {\sc Scamp} \citep{2006ASPC..351..112B}. 

The photometric zero-points were derived using standard fields obtained the same night whenever available, or all-sky catalogues such as 2MASS \citep{2006AJ....131.1163S} in J, H and Ks, SDSS in $g,r,i,z$ and IPHAS \citep{IPHAS} in H$\alpha$. The zeropoint rms were typically in the range between 0.02 and 0.12~mag, and were added quadratically to the measurement uncertainty. No attempt was made to calibrate the non-standard VR filters used with the Mosaic and SuprimeCam cameras and the He~I narrow band filter obtained with the WFC. 

\begin{figure}
   \centering
   \includegraphics[width=0.45\textwidth]{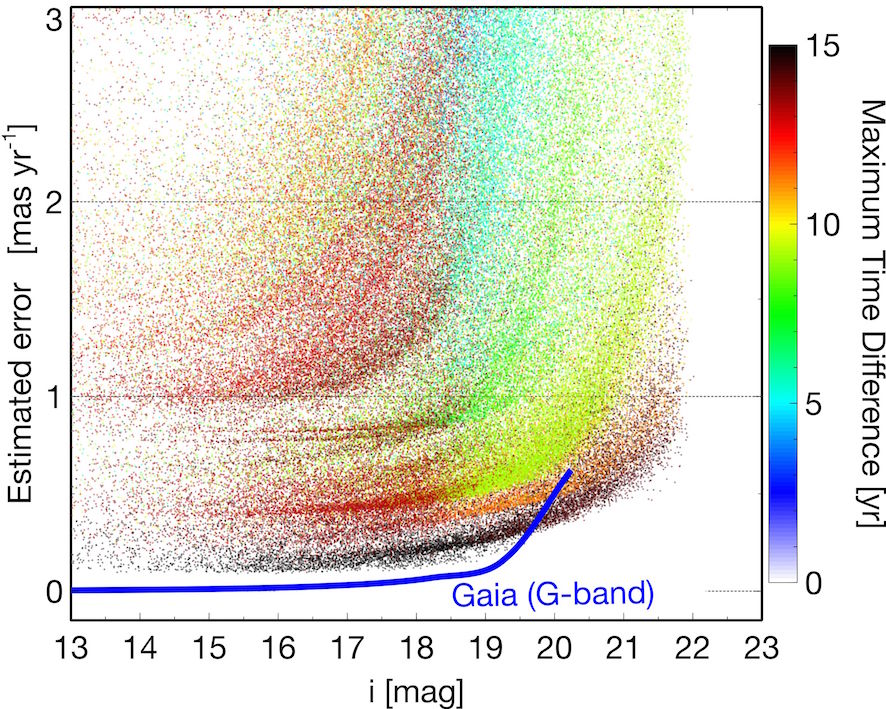}
      \caption{Estimated error on the proper motion as a function of $i$-band magnitude. The color scale represents the maximum time difference used for each measurement. The prediction for {\it Gaia} is over-plotted for comparison.}
         \label{fig:error}
   \end{figure}

Proper motions for all the sources were computed using the method described in \citet{2013A&A...554A.101B}. The final residuals on the astrometric solution show a dispersion of the order of 11$\sim$12~mas. Figure~\ref{fig:error} shows the estimated error on the proper motion as a function of $i$-band apparent magnitude. We note that the proper motion measurements are not tied to an absolute reference system such as the ICRS (International Celestial Reference System). They are nevertheless expected to be close to the ICRS since extragalactic sources and background stars largely dominate the sample and display very little motion with respect to the ICRS.

A final master catalog of all sources (338\,892 in total) was then produced including the measured photometry in the $grizJHKs$ filters. A total of 194\,452 sources (approximately half) were detected at more than two epochs and their proper motion are also included. Since the main goal of the survey focuses on the study of M35, only objects with  $|\mu_{\alpha} \cos \delta|\le$100~mas yr${-1}$ and $|\mu_{\delta}|\le$100~mas yr$^{-1}$ were considered for the membership analysis.

\section{Membership}
We use the method described in \citet{2014A&A...563A..45S} to derive membership probabilities of all the sources. Briefly, the method uses a training set -- in this case the sample of candidate members from \citet{2001ApJ...546.1006B} -- to identify the features that most efficiently separate the cluster's population from the field and background population among all combinations of proper motion, luminosities and colors available. In the case of M35, the best set was found to include the proper motions, the $z$, $H$, $Ks$ luminosities and the $(r-z)$, $(r-i)$ and $(J-H)$ colors. Once the best set of colors, luminosities and proper motion is selected, a model of the cluster's locus defined by the training set is fitted in the corresponding multi-dimensional space. A mixture of Gaussians is used for the proper motion diagram and principal curves for the cluster's sequence in all color-magnitude and color-color diagrams. Membership probabilities are then computed using this first model. The highest probability members are selected as new training set, and the models and the membership probabilities are recomputed until they converge.

\begin{figure}
   \centering
   \includegraphics[width=0.45\textwidth]{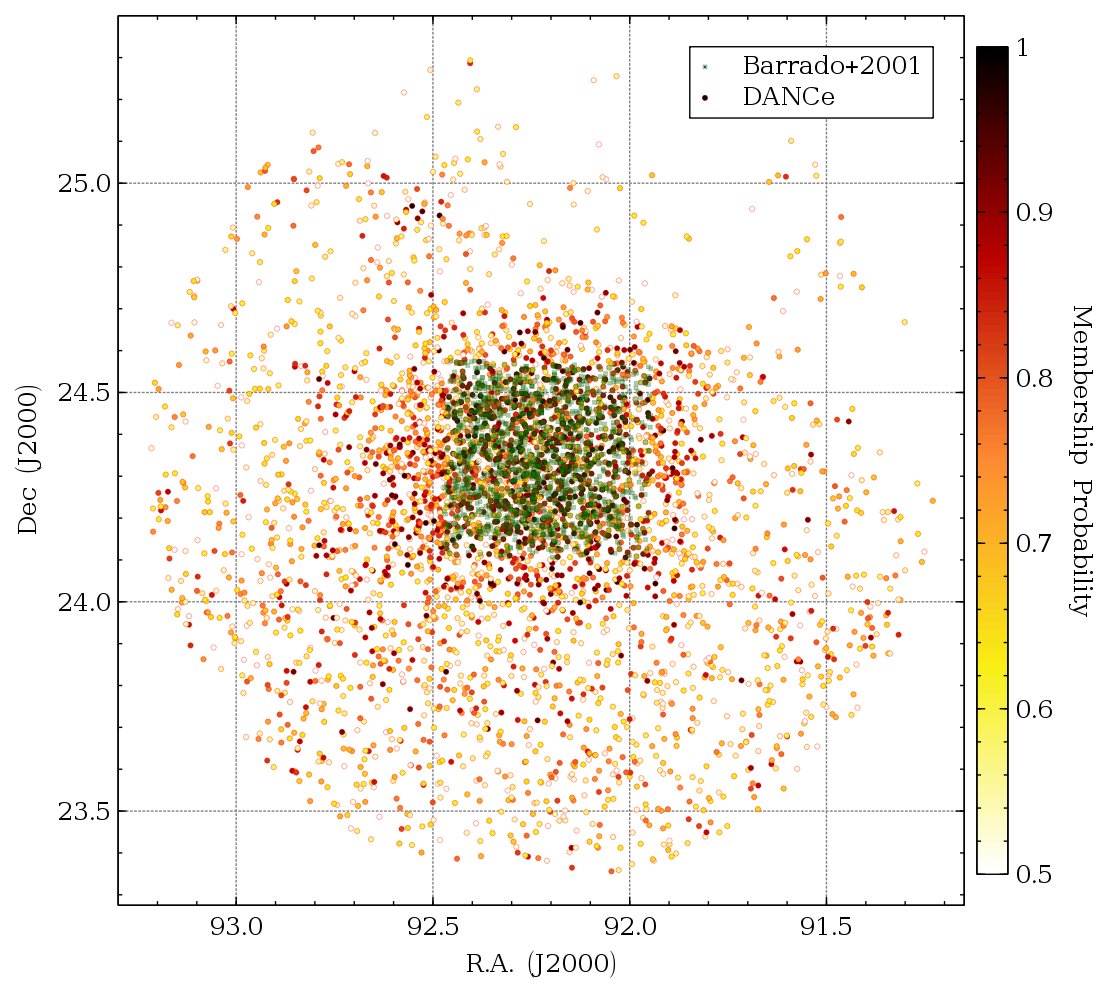}
      \caption{Spatial distribution of sources used as training set \citep[green squares, from ][]{2001ApJ...546.1006B} and of the 4\,349 sources of the final sample with membership probability greater than 0.5. Background image: 2MASS J-band.}
         \label{fig:map}
   \end{figure}

This method was originally developed and optimized for clusters like the Pleiades, i.e clusters with a relatively large proper motion and photometric sequence clearly separated from the field and background sequences in many color-magnitude diagrams. Unfortunately M35 is a worst-case scenario for the method. Its proper motion is relatively small and its sequence largely overlaps with the field and background sequence in almost all color-magnitude and color-color diagrams. To limit the contamination and enhance the contrast between cluster members and field and background sources, we chose to restrict the analysis to a radius of 1\degr\, around the cluster's center located at $\alpha$=06h08m54.0s, and $\delta$=+24\degr20\arcmin00\arcsec (J2000).
Figure~\ref{fig:map} shows their spatial distribution. It shows that the north-western part of the survey was shallower than the rest of the survey, as fewer members were selected. It also shows that a number of high probability members have $(r-i)$ or $(i-Ks)$ color apparently inconsistent with the cluster's sequence. Their high probability is due to their proximity to the isochrone in all other color-magnitude and color-color diagrams as well as to the cluster's locus in the proper motion diagram. These sources -- probably suffering from inaccurate photometry in one of these bands -- illustrate the robustness of the selection method based on a high dimensional space.

Table~\ref{tab:catalog} gives the astrometry, photometry and membership probability for the 338\,892 unique sources detected in the corresponding area. Note that only the sources with a proper motion measurement have a membership probability (194\,452 in total), as the availability of a proper motion measurement was a necessary condition in our membership analysis. Using sources with membership probability greater or equal to 0.9, we derive a median proper motion for the cluster of ($\mu_{\alpha} \cos \delta, \mu_{\delta}$)=(1.99, -0.23)$\pm$(0.82,0.71)~mas yr${-1}$, where the uncertainties correspond to the scaled median absolute deviation. 

   \begin{figure*}
   \centering
   \includegraphics[width=0.95\textwidth]{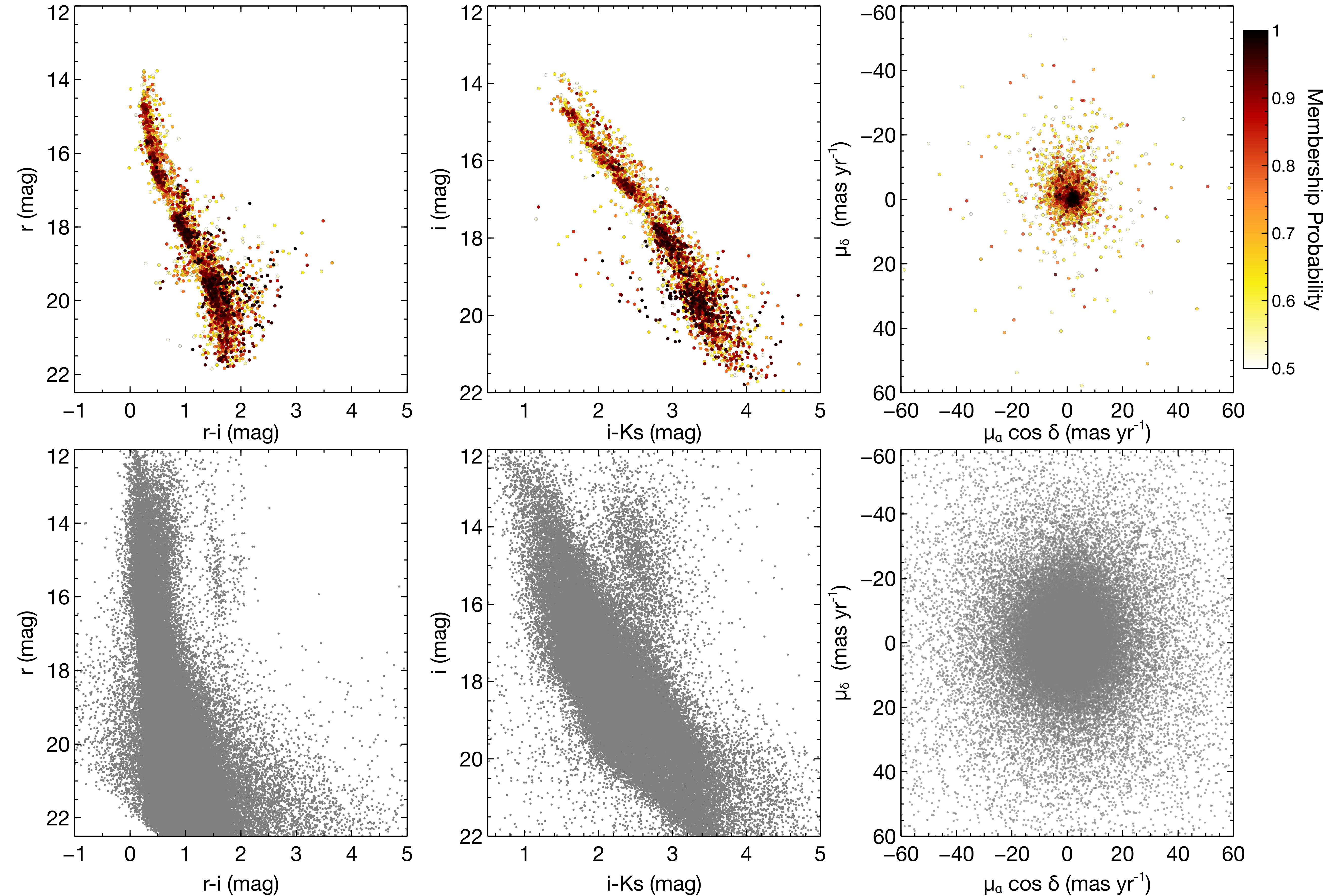}
      \caption{Color magnitude and proper motion diagrams of the 194\,452 sources of the sample detected at more than two epochs. Left: $r$ vs $r-i$. Middle: $i$ vs $i-Ks$. Right: proper motion diagram. Sources with a membership probability greater or equal to 0.5 are represented in the top panels with a color scale proportional to their probability. The rest of the sources is represented in the bottom panels.}
         \label{fig:cmd}
   \end{figure*}

\begin{figure}
   \centering
   \includegraphics[width=0.45\textwidth]{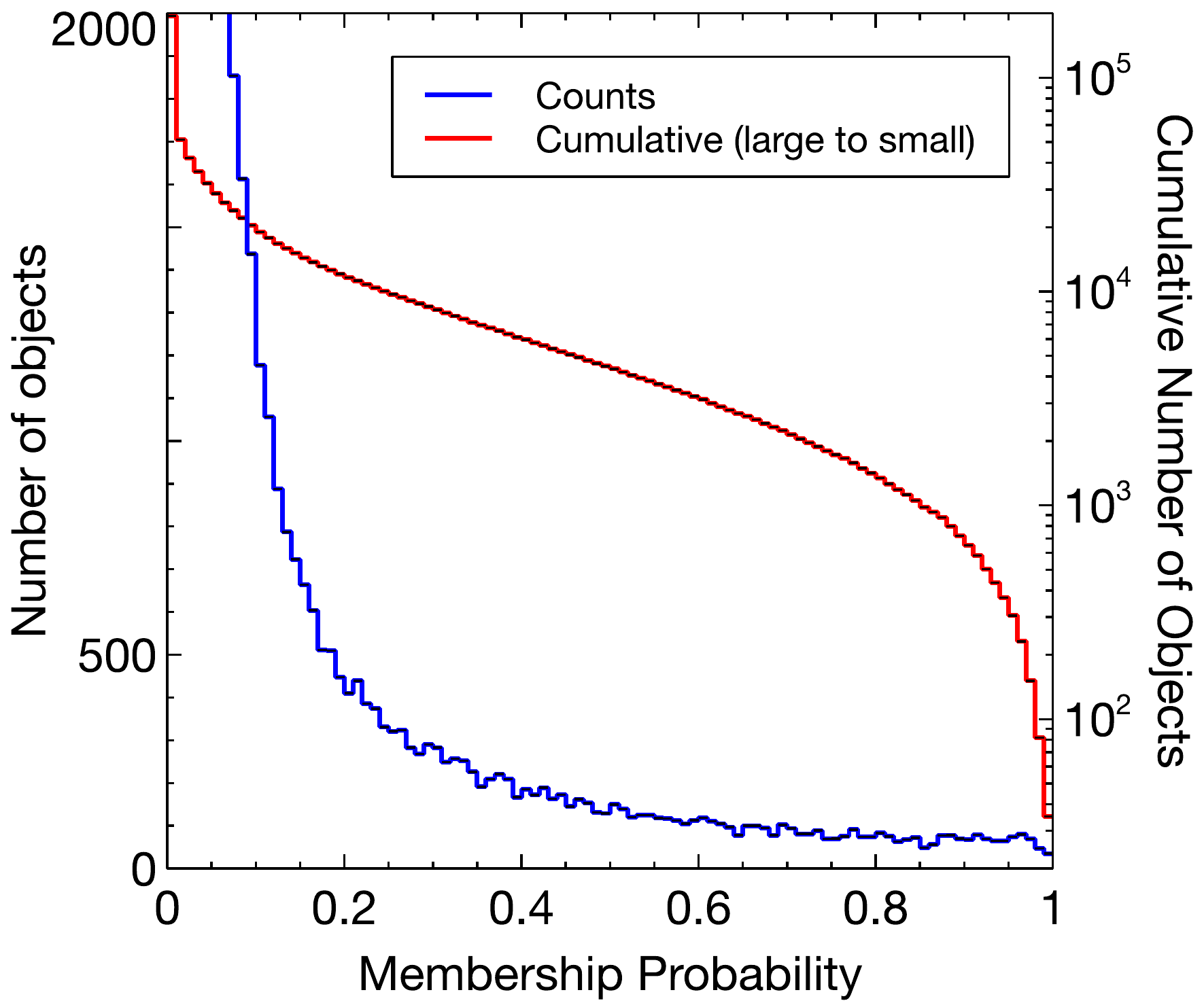}
      \caption{Distribution (blue, left scale) and cumulative distribution (red, computed from large to small values, right scale) of membership probabilities for the 194\,452 sources of the sample}
         \label{fig:proba}
   \end{figure}

Figure~\ref{fig:proba} shows the membership  distribution and cumulative distribution (computed from large to small) for all the sources in the catalog. The vast majority of the sources has a membership probability of $\approx$0.0. A monotonously decreasing distribution fills the probability range between 0 and 1.0, as the result of the lose constraints on the photometric and astrometric properties of the cluster. 

\begin{figure*}
   \centering
   \includegraphics[width=0.95\textwidth]{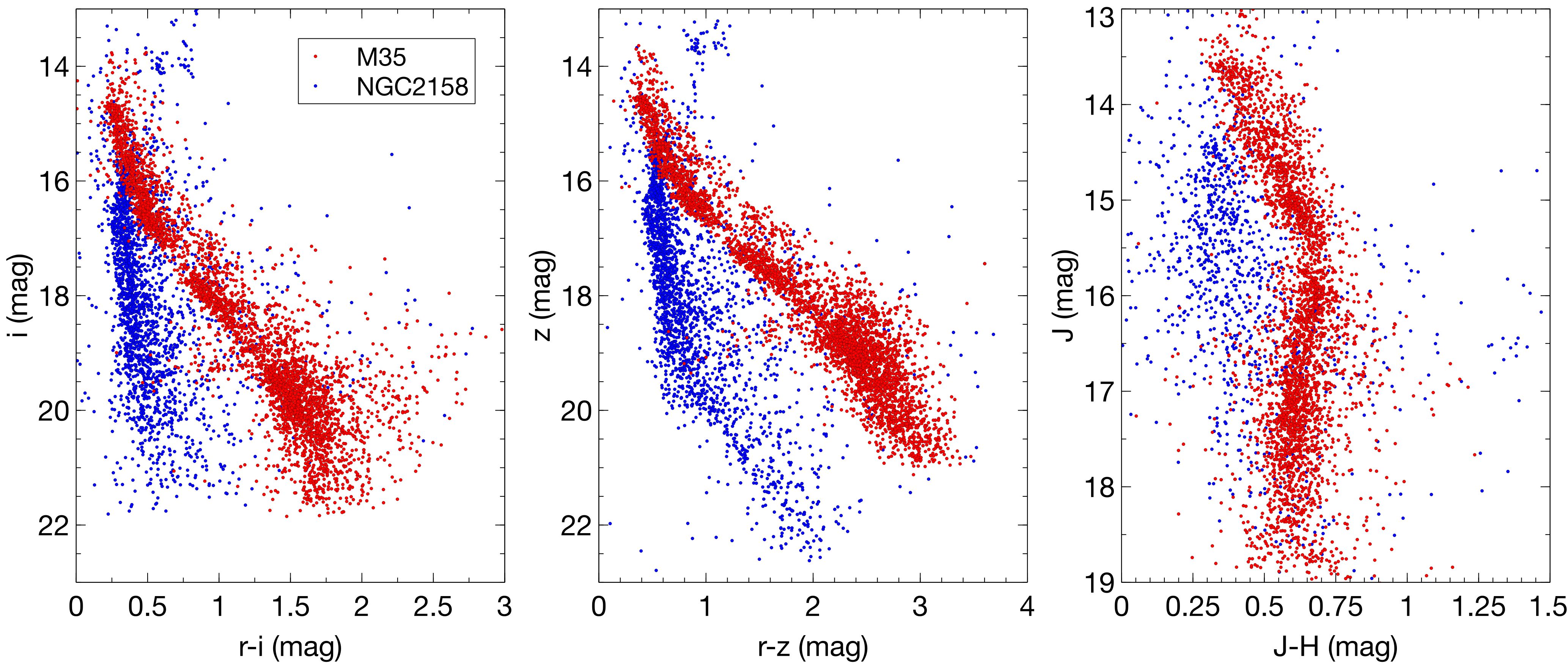}
      \caption{Color-magnitude diagrams of the M35 high probability members ($\ge$0.5. red dots) and of all the sources within 3\arcmin\, of NGC~2158 center (blue dots).}
         \label{fig:ngc2158}
   \end{figure*}

NGC~2158 is an intermediate age (1$\sim$2~Gyr) cluster at a distance of $\approx$3\,600~pc \citep{2002MNRAS.332..705C} located at a projected distance of only 25\arcmin\, from M35. It is therefore included in the area covered by the present study and possibly contaminates the membership analysis. To estimate the level of contamination due to NGC~2158 members, we select all the sources located within 3\arcmin\, of its center. Given the distance and projected density profile of NGC~2158, these sources are expected to be in their vast majority members of NGC~2158. Figure~\ref{fig:ngc2158} shows the sequence formed by these sources in various color-magnitude diagrams made of colors and luminosities used for the membership analysis presented here, and compares it to the sequence formed by objects with a probability of membership to M35 greater than 0.5 according to our analysis. The two sequences overlap mostly at the bright end, and contamination must occur mostly above $i\lesssim$17~mag. We find that only 27 sources with a probability of membership to M35 greater than 0.5 are located within 3\arcmin\, of NGC~2158, and 53 within 5\arcmin. Given the projected size of NGC~2158 core, little contamination is expected beyond 5\arcmin, and the contamination by NGC~2158 members must not reach more than 1$\sim$2\% at most.

\section{Conclusions}
We have processed and analyzed a total of 4\,867 wide-field optical and near-infrared images of M35 obtained from various public astronomical archives. The images cover 18 years of time baseline and typically reach a depth of 21$\approx$22~mag in the optical and $\approx$19~mag in the near-infrared, while saturation arises around 12$\sim$13~mag at the bright end. This dataset allows us to derive multi-wavelength photometry from the optical to the near-infrared for 338\,892 sources, and proper motions for 194\,452 of them. We use these measurements to derive the membership probability of all the sources with a proper motion measurement. We obtain:
\begin{itemize}
\item 4\,349 candidate members with probability greater than 50\%
\item 1\,726 candidate members with probability greater than 75\%
\item 305 candidate members with probability greater than 95\%
\end{itemize}
These numbers are consistent with the census presented in \citet{2001ApJ...546.1006B} over a smaller area. Contamination is expected to be significant as the locus of the cluster in the proper motion, color-magnitude and color-color diagrams overlaps largely with the field and background locus. This new study nevertheless provides a coherent and quantitative membership analysis of M35 based on a large fraction of the best ground-based datasets obtained over the past 18 years. 

\begin{acknowledgements}
 H. Bouy is funded by the the Ram\'on y Cajal fellowship program number RYC-2009-04497. This research has been funded by Spanish grants AYA2012-38897-C02-01. E. Moraux ackowledges funding from the Agence Nationale pour la Recherche program ANR 2010 JCJC 0501 1 ``DESC (Dynamical Evolution of Stellar Clusters)''. A. Ribas is funded by ESAC Science Operations Division research funds with code SC 1300016149. Based on observations obtained with MegaPrime/MegaCam, a joint project of {\it CFHT} and CEA/DAPNIA, at the Canada-France-Hawaii Telescope (CFHT) which is operated by the National Research Council (NRC) of Canada, the Institut National des Science de l'Univers of the Centre National de la Recherche Scientifique (CNRS) of France, and the University of Hawaii. Based on observations obtained with WIRCam, a joint project of CFHT, Taiwan, Korea, Canada, France, at the Canada-France-Hawaii Telescope (CFHT) which is operated by the National Research Council (NRC) of Canada, the Institute National des Sciences de l'Univers of the Centre National de la Recherche Scientifique of France, and the University of Hawaii. This paper makes use of data obtained from the Isaac Newton Group Archive which is maintained as part of the CASU Astronomical Data Centre at the Institute of Astronomy, Cambridge. The data was made publically available through the Isaac Newton Group's Wide Field Camera Survey Programme. The Isaac Newton Telescope is operated on the island of La Palma by the Isaac Newton Group in the Spanish Observatorio del Roque de los Muchachos of the Instituto de Astrofísica de Canarias. This research used the facilities of the Canadian Astronomy Data Centre operated by the National Research Council of Canada with the support of the Canadian Space Agency. This research draws upon data distributed by the NOAO Science Archive. NOAO is operated by the Association of Universities for Research in Astronomy (AURA) under cooperative agreement with the National Science Foundation. This publication makes use of data products from the Two Micron All Sky Survey, which is a joint project of the University of Massachusetts and the Infrared Processing and Analysis Center/California Institute of Technology, funded by the National Aeronautics and Space Administration and the National Science Foundation. This work is based in part on data obtained as part of the UKIRT Infrared Deep Sky Survey.  This research has made use of the VizieR and Aladin images and catalogue access tools and of the SIMBAD database, operated at CDS, Strasbourg, France. This research made use of data from the SDSS survey. Funding for the creation and distribution of the SDSS Archive has been provided by the Alfred P. Sloan Foundation, the Participating Institutions, the National Aeronautics and Space Administration, the National Science Foundation, the U.S. Department of Energy, the Japanese Monbukagakusho, and the Max Planck Society. The SDSS Web site is http://www.sdss.org/. The Participating Institutions are The University of Chicago, Fermilab, the Institute for Advanced Study, the Japan Participation Group, The Johns Hopkins University, the Max-Planck-Institute for Astronomy (MPIA), the Max-Planck-Institute for Astrophysics (MPA), New Mexico State University, Princeton University, the United States Naval Observatory, and the University of Washington. Based on data collected at Subaru Telescope and Kiso observatory (University of Tokyo) and obtained from the SMOKA, which is operated by the Astronomy Data Center, National Astronomical Observatory of Japan.
\end{acknowledgements}

\bibliographystyle{aa}
\bibliography{mybiblio}

\end{document}